# Petahertz Spintronics

Florian Siegrist[1,5], Julia A. Gessner[1,5], Marcus Ossiander[1], Christian Denker[2], Yi-Ping Chang[1], Malte C. Schröder[1], Alexander Guggenmos[1,5], Yang Cui[5], Jakob Walowski[2], Ulrike Martens[2], J. K. Dewhurst[4], Ulf Kleineberg[1,5], Markus Münzenberg[2], Sangeeta Sharma[3], Martin Schultze[1,5,*]

[1]Max-Planck-Institute of Quantum Optics, Hans-Kopfermann-Straße 1, 85748 Garching
[2]Institut für Physik, Universität Greifswald, Felix-Hausdorff-Straße 6, 17489 Greifswald
[3]Max Born Institute for Nonlinear Optics and Short Pulse Spectroscopy, Max-Born-Straße 2A, 12489 Berlin
[4]Max-Planck-Institute of Microstructure Physics, Weinberg 2, 06120 Halle (Saale)
[5]Fakultät für Physik, Ludwig-Maximilians-Universität München, Am Coulombwall 1, 85748 Garching, Germany
*Correspondence to: martin.schultze@mpq.mpg.de

**The enigmatic coupling between electronic and magnetic phenomena was one of the riddles propelling the development of modern electromagnetism[1]. Today, the fully controlled electric field evolution of ultrashort laser pulses permits the direct and ultrafast control of *electronic* properties of matter and is the cornerstone of light-wave electronics[2–7]. In sharp contrast, because there is no first order interaction between light and spins, the *magnetic* properties of matter can only be affected indirectly on the much slower tens-of-femtosecond timescale in a sequence of optical excitation followed by the rearrangement of the spin structure[8–16].**

**Here we record an orders of magnitude faster magnetic switching with sub-femtosecond response time by initiating optical excitations with near-single-cycle laser pulses in a ferromagnetic layer stack. The unfolding dynamics are tracked in real-time by a novel attosecond time-resolved magnetic circular dichroism (atto-MCD) detection scheme revealing optically induced spin and orbital momentum transfer (OISTR) in synchrony with light field driven charge relocation[17]. In tandem with *ab-initio* quantum dynamical modelling, we show how this mechanism provides simultaneous control over electronic and magnetic properties that are at the heart of spintronic functionality. This first incarnation of attomagnetism observes light field coherent control of spin-dynamics in the initial non-dissipative temporal regime and paves the way towards coherent spintronic applications with Petahertz clock rates.**

While matter reacts to optical excitations by a virtually instantaneous and direct response of electrons to the electric field oscillations, ultrafast magnetic switching is intrinsically slower due to the need of mediating processes linking optical excitations to spin dynamics[8,9,13,14]. Femtosecond laser pulses can modify magnetic properties at response times of several tens of femtoseconds or longer $(50 - 500 \text{ x } 10^{-15}\text{s})$[15,16], the prospect of optical control of ferromagnetic bits has been discussed[18] and circularly polarized attosecond light pulses are proposed to generate ultrafast magnetic field bursts as new tools for ultrafast magneto-optics[19]. However, direct experimental evidence of the ability to link the response of a spin system to the quasi-instantaneous opto-electronic response calls for attosecond temporal resolution and is missing to date.

A route to the novel regime of attosecond magnetism has been worked out theoretically and proposes the use of ferromagnetic alloys or layer stacks where optical excitations result in the local displacement of charge carriers between different atomic species or across layer interfaces, in analogy to shift currents discussed in compound semiconductors[20,21]. In such a scenario, dubbed OISTR, the spatially dislodged electron wave carries its spin away from the atomic species at which its ground state resides[17]. Consequently, the resulting coherent spin transfer is linked directly to the temporal evolution of the optical excitation field and modifies the spin moment of the magnetic layers extended over macroscopic dimensions defined by the illuminated area.

Critical for advancing towards attosecond magnetism is an experimental scheme with the ability to control and detect the sub-optical-cycle evolution of electronic excitations driven by ultrafast electric fields, combined with simultaneous observation of modifications of the magnetic moment of individual sample constituents. Here we achieve this by linking attosecond transient absorption detection (atto-XAS), which is sensitive to the temporal evolution of electronic excitations[2–7], with simultaneous atto-MCD sketched in Fig. 1, which is probing the local magnetic moment.

In our experiment, ultrafast near-infrared (NIR) few-cycle laser pulses serve as temporally well confined trigger of electronic excitations in nickel. The metal exhibits broadband resonant absorption properties (see density of states DOS , Fig. 1 **c**) and the strong electric field of the laser pulse accelerates electrons around the Fermi energy in the band structure or promotes them to states in higher energy bands. Similarly, for nickel sandwiched between platinum layers the optical excitation is a virtually instantaneous response to the external electric field oscillations. However, the adjacent heavy metal ad-layers open an additional excitation pathway where an electron is coherently driven across the material interface from states around the Fermi energy of nickel into vacant platinum states (Fig. 1 **c**, **d**).

The resulting depletion of majority spins oriented by placement of the samples in a static external *B*-field well beyond magnetic saturation ($B_{ext} = 50$ mT) leads to a macroscopic reduction of the nickel layer magnetic moment. We employ atto-MCD as illustrated in Fig. 1 for the first real time study of this light field induced coherent modification of a materials' magnetic moment.

Linearly polarized attosecond pulses are generated by the same ultrafast laser pulses used to pump the electronic system of the samples via high-harmonic-generation and subsequent spectral selection[22–24]. To implement a time resolved variant of x-ray magnetic circular dichroism detection[25], we developed a grazing incidence, multi reflection phase-retarder[26,27] (Fig. 1 **a**) optimized for photon energies in the extreme ultraviolet (XUV) regime. Rotation of the phase-retarder around the beam propagation axis allows either to transmit the initially linear polarized light or to create circularly polarized attosecond pulses (see Methods).

In our experiment, the resulting circularly polarized attosecond pulses cover XUV photon energies around the Nickel $M_{2,3}$-transition (Fig. 1) and last ~20% of the half-cycle duration of the optical pump field ($\tau_{XUV} = 310$ as) as confirmed by attosecond streak camera detection.

In a transient absorption scheme, these pulses permit to measure the dynamics of photo-excited electrons and modifications of the spin structure. Final state interactions between electrons photo-injected into conduction band (CB) states and electrons promoted into the CB by an XUV photon turn the energy resolved XUV absorption into a sensitive probe for pump laser induced population transfer between electronic states around the Fermi edge of Nickel[28].

Recording the transmitted XUV spectral intensities $I^{+/-}$ for two opposite magnetization directions oriented along the sample surface (or equivalently for two orthogonal helicities of the attosecond pulses) yields the magnetic dichroism contrast $\Delta_M = I^+ - I^-$ (atto-MCD, see Fig. 1 **b**, and Methods). Detection of $\Delta_M$ renders the method sensitive to element specific local magnetic moments and allows the simultaneous tracking of sub-femtosecond time-resolved modifications of both the electronic and magnetic implications of strong field induced optical pumping.

Fig. 2 shows the recorded atto-XAS and atto-MCD transients. Ultrashort NIR waveforms carried at a central wavelength of $\lambda_{NIR} = 800$ nm, set to a peak intensity of $I_{NIR} = 4 \times 10^{12}\,\text{W}/\text{cm}^2$ and full-width-at-half-intensity-maximum duration of $\tau_{NIR} = 4$ fs are used to excite electrons in nickel (panel **b**, sample **I**, for details on sample preparation and static characterization see Methods). Owing to the attosecond temporal resolution of the measurement, in synchrony with the electric field oscillations (simulated waveform depicted as green line, top of panel **a**) a repetitive decrease of the XUV absorbance is observed (red line)[3]. The attosecond pulse experiences this increased absorption (energy interval $\Delta E = 65.5 - 66.5$ eV) as a result of reduced Pauli blocking in electronic states that gradually are liberated by the pump field induced promotion of carriers into states above the Fermi energy. The stepwise behavior is indicative of the transition rate peaking at the field crests of the light waveform and the number of excited carriers increasing in synchrony with the half-cycle oscillations of the NIR pump laser pulses, reminiscent of interband tunneling[3,5,29].

Inspection of the simultaneously recorded atto-MCD contrast $\Delta_M$ (blue line) yields no observable change within the sampled time interval, indicating the conservation of magnetic moment in the ferromagnetic nickel layer for the first $\sim 10$ fs during and after electronic excitation.

In striking contrast, panel **a** shows the magnetic moment of nickel sandwiched between platinum layers (sample **II**, see Methods) under otherwise identical experimental conditions to respond as fast as the electronic system to the optical excitation. Platinum surrounding nickel acts as efficient spin absorber for optical excitations transferring charge across the interfaces. In these samples, ultrafast changes of $\Delta_M$ (blue line) $> 40\%$ of the initial value are immediately discernible during the light field oscillations that cause optical excitation, attesting to the dominant role of OISTR in the loss of magnetic moment. The reduction acts out with a time constant of $\tau^{coherent} = 4.5$ fs ($1/e$), concurrent with the field induced excitation and identical to the response function computed from the laser pulse duration. This synchrony is direct evidence for coherent, sub-femtosecond magnetization control extended over the macroscopic dimensions probed by the attosecond pulse spot size $\emptyset_{XUV} \approx 40 \mu m$.

To shed light on the coherent electronic processes in the presence of the oscillating light field and its influence on initial stages of ultrafast spin dynamics, we turn to theory. Fig. 3 compares the experimental data with the theoretical results obtained using a state-of-the-art, full quantum description of the dynamics of the non-equilibrium state of matter by means of time-dependent density functional theory (TD-DFT), which accounts for the coupled dynamics of both, charge and spins. The theory identifies the two major phenomena that contribute to the dynamics in the early times (during and immediately after pumping) to be the flow of charge and spin currents across the interface and spin-orbit-mediated spin-flips. The net result of which is coherent momentum transfer (CMT) in space and time across the interface by redistribution of the interlinked spin **S**(t) and orbital angular momentum **L**(t).

The combination of TD-DFT and atto-MCD, both capable of tracking ultrafast charge and spin migration that outpace the slower lattice dynamics, is therefore a powerful framework to investigate how different mechanisms conspire to turn the initially coherent optical excitation into a macroscopic loss of magnetic moment. Fig. 3**a** shows the results of atto-MCD recording the magnetization dynamics in Ni/Pt sandwich samples pumped at an intensity of $I_{NIR} = 2 \times 10^{12}$ W/cm$^2$ (green line w. circles) in comparison with the results of *ab-initio* computations with the sample geometry and experimental laser pulse parameters as input. Excellent quantitative agreement between experimentally observed $\Delta_M$ and the results of the full calculation (including spin-orbit coupling) of the time evolution of the relative magnetic moment $M(t)/M(0)$ testifies for the validity of the theoretical approach and confirms that the experimental observable is a direct measure of the layer magnetization. Interestingly, excluding spin-orbit coupling from the theoretical calculations (dark red line) causes a leveling off from the predicted magnetization dynamics at 20 fs, which is not mirrored in the experimental data. Since the only process that can cause a demagnetization of Ni apart from spin-orbit coupling is the flow of electrons (carrying their spins) across the interface, these first 20 fs mark the hitherto unexplored time interval in which OISTR-mediated CMT dictates the physics of spin-dynamics.

Our data indicate the presence of a second time scale of slower demagnetization (20 fs $< t <$ 50 fs) where theory identifies spin-orbit-mediated spin-flips to dominate the evolution of the magnetic moment. Later after optical excitation, ($t > 50$ fs, panel **b**) the sample fully demagnetizes with a timescale of $\tau^{incoherent} = 112 \pm 6$fs which represents the transition to stochastic dynamics rooted in magnetically correlated behavior and is in agreement with previous experiments reporting the decay of magnetic moment in the range of $80 - 400$ fs depending on system dependent quench rates[10,12].

We observed light-wave induced coherent transfer of spin and orbital angular momentum in space and time caused by the interplay of the few-cycle optical excitation and the spin-orbit interaction in magnetic/non-magnetic multilayers. This excursion into the unexplored territory of ultrafast ($t < 20$ fs) all-optical control of spin dynamics and macroscopic magnetic moments by and in synchrony with the oscillations of ultrafast electric fields opens the door to the novel regime of attosecond magnetism and is a benchmark for the design of future coherent magnetic control protocols. Our *ab initio* theory predicts that in alloys or suitably chosen multilayer systems the same mechanism can be tailored to cause a local and ultrafast increase in magnetic moment to sustain reversible optical switching of magnetic moments bringing spintronics to the attosecond regime. Finally, the observed sub-femtosecond magnetization control via optically induced spin transfer, promises a new class of ultrafast spintronic applications, and suggests the feasibility of light wave gated coherent spin transistors mimicking spintronic functionality orders of magnitude faster than applicable today.

**Methods:**

**Theory:**

Computations rely on the Runge-Gross theorem[30] establishing that the time-dependent external potential is a unique functional of the time dependent density, given the initial state. Based on this theorem, a system of non-interacting particles can be chosen such that the density of this non-interacting system is equal to that of the interacting system for all times. The wave function is represented as a Slater determinant of single-particle Kohn-Sham (KS) orbitals. A fully non-collinear spin-dependent version of this theorem entails that these KS orbitals are two-component Pauli spinors determined by the equations:

$$i\frac{\partial \psi_j(r,t)}{\partial t} = \left[\frac{1}{2}\left(-i\nabla + \frac{1}{c}A_{ext}(t)\right)^2 + v_s(r,t) + \frac{1}{2c}\sigma \cdot B_s(r,t) + \frac{1}{2c}\sigma \cdot (\nabla v_s(r,t) \times -i\nabla)\right]\psi_j(r,t) \quad (1)$$

where the first term is the kinetic term and responsible for the flow of current across the interface[31], $A_{ext}(t)$ is a vector potential representing the applied laser field, and $\sigma$ are the Pauli matrices. The KS effective potential $v_s(r,t) = v_{ext}(r,t) + v_H(r,t) + v_{XC}(r,t)$ is decomposed into the external potential $v_s(r,t)$, the classical electrostatic Hartree potential $v_H(r,t)$ and the exchange-correlation (XC) potential $v_{XC}(r,t)$. Similarly, the KS magnetic field is written as $B_s(r,t) = B_{ext}(r,t) + B_H(r,t) + B_{XC}(r,t)$ where $B_{ext}(r,t)$ is the magnetic field of the applied laser pulse plus possibly an additional magnetic field and $B_{XC}(r,t)$ is the XC magnetic field. The final term is the spin-orbit coupling term, presence of which ensures that the total spin angular momentum, **S**, is not a good quantum number and hence spin-flip excitations can lead to loss in **S**. Eq. 1 is solved for the electronic system alone in dipole approximation and by using adiabatic local spin density approximation[32] for the XC fields.

Dynamics of the nuclear degrees of freedom and radiative effects described by simultaneously time-propagating Maxwell's equations, are not included in the present work limiting the predictive power of the method to the first ~100 fs during and after pumping.

Calculations of a magneto-optical function for NiPt layer (see inset Fig. 3) were performed by a 3 step process: (i) The ground-state of NiPt multilayers was determined using DFT, (ii) a fully spin-polarized GW calculation[33] was performed to determine the correct position and width of Ni 3p states and (iii) finally the response function was calculated on top of GW-corrected Kohn-sham ground-state. This response function is calculated within linear response TD-DFT in which the excitonic effects[34] and local field effects can be easily included. As a merit of this treatment, no experimental parameter was needed to determine the accurate magneto-optical functions.

Computational details:

A fully non-collinear version of TDDFT as implemented within the Elk code[35] is used for all calculations. All the implementations are done using the state-of-the art full potential linearized augmented plane wave (LAPW) method.

A regular mesh in $\mathbf{k}$-space of $8 \times 8 \times 1$ is used and a time step of $\Delta t = 2$ as is employed for the time-propagation algorithm. To mimic experimental resolution, a Gaussian energy broadening is applied with spectral width of 0.027 eV. The laser pulse used in the present work is linearly polarized (out of plane polarization) with central frequency of 1.55 eV, full-width-at-half-maximum duration 8fs and fluence of 5.4 mJ/cm².

**Experiment:**

The experiments are carried out with a phase-stabilized few-cycle near-infrared laser as driver (FemtoPower). It delivers pulses with 4fs, 0.5mJ at a repetition rate of 4kHz. These pulses are focused into a neon-filled ceramics target, generating high-harmonic radiation with a cut-off energy around 70 eV. A Mach-Zehnder type interferometer is used to introduce a delay between the NIR and the XUV pulses. We use a 150nm Al filter to block the driving NIR laser pulse, while providing a constant transmission at the energies of interest. Isolated attosecond pulses are achieved by spectrally filtering the cut-off regime with a Si/B4C multilayer mirror with a reflection of 15% @66eV and a full-width-at-half-maximum bandwidth of 8eV under and angle of incidence of 45°. Subsequently, the multi reflection phase-shifter changes the polarization from linear to elliptical. Ellipticity of ε>0.75 is achieved with transmittance of >25% and without introducing significant wavelength dispersion,

Four Mo/B4C multilayer mirrors are mounted under an angle of 78° with respect to the surface normal, a setting optimized for high broadband transmission and maximized ellipticity.

To maximize the atto-MCD contrast while maintaining sufficient XUV transmission the samples are mounted under an angle of ~35° between the propagation direction of the laser and the surface normal. The surface normal and the magnetization direction of the sample span a plane parallel to the propagation direction of the laser field. We use a gold coated grating with 2105 lines/mm in reflection (Jobin-Yvon) with a 200 µm entrance slit as a spectrometer.

For every delay step, we measured the transmitted spectrum for both magnetization directions of the sample and then scanned the delay over the region of interested. To check and compensate for long timescale drifts in the XUV spectrum we took spectra of the transmitted XUV in the absence of NIR laser light before and after every scan. to ensures that no irreversible changes were done to the multilayer or the supporting polycrystalline Silicon substrate. Typical integration times were on the order of 10-15s for every spectrum, leading to a measurement duration of 30-60 minutes.

We observe the OISTR effect for pump light intensities ranging between $I_{NIR} = 1.5 - 4 \times 10^{12} \text{ W}/\text{cm}^2$. At lower intensities, only the slow demagnetization component depicted in Fig. 3, panel **b** is observed. At intensities $I_{NIR} > 5 \times 10^{12} \text{ W}/\text{cm}^2$ the time span between laser pulses (250 µs) is found to be insufficient for the magnetic system to return to the unperturbed state.

**Data Evaluation:**

For the energy calibration of our spectrum we used the pronounced aluminum $L_{2,3}$-edges at 72.7 eV and the Nickel $M_3$-edge at 66.2 eV. The grating equation $n * \lambda = d * (\sin \theta_i - \sin \theta_f)$ gives a relation between the position of a spectral feature on the XUV sensitive camera (Princeton Instrument PIXIS) and the photon energy. We checked for the influence of the non-homogeneous diffraction and reflection efficiency of the grating and found it to be negligible within the range of photon energies relevant in this study. We applied an equal weight sliding average to the spectrograms along the energy ($\Delta E = 50$ meV) as well as along the delay dimension ($\Delta t = 0.5$ fs). We determine the magnetic dichroism contrast according to $\Delta_M = I^+ - I^-$. Application of alternative definitions of the contrast or computing the magnetic dichroism asymmetry has no influence on the transient signals.

**Sample preparation and magnetic characterization:**

A number of Ni and Ni/Pt multilayer films were grown on Si membranes and optimized to grow in fcc structure. The 8.4 nm Ni film and Pt(2)/[Ni(4)/Pt(2)]×2/Ni(2nm) multilayer were deposited on 200 nm thick polycrystalline Silicon membranes (Norcada) at room temperature by electron beam evaporation under ultrahigh vacuum (UHV) conditions with a base pressure less than $1 \times 10^{-9}$ mbar with high purity source materials (99.95%). The individual layer thicknesses in the [Ni(4)/Pt(2)]×2 stack were optimized to increase the number of active interfaces for OISTR while maintaining sufficient XUV transmittance and providing maximized atto-MCD contrast. The sample was protected by an inert 2 nm Pt capping layer and an ultrathin Ni seed layer (nonmagnetic at room temperature) was grown on the substrate to provide identical surrounding to all Pt layers and ensure uniform growth of the layer stack. Both materials grow in fcc structure. The deposition rate was monitored by a calibrated standard quartz crystal microbalance and adjusted to 0.02 nm/s. Special care was taken for low tension mounting to protect the membranes from any mechanical stress. The in-plane magnetic properties were characterized by a standard longitudinal Magneto-Optical Kerr Effect (MOKE) setup with a 3 mW HeNe cw-laser, a photoelastic modulator (PEM) and Glan-laser prisms in longitudinal geometry. Measurements on both, the frame and the samples membrane window, yield similar ferromagnetic properties. For both types of samples, we found coercive fields of a few mT and a remanence around 60-70% of the saturation magnetization. MOKE measurements confirm that the Ni seed layer does not magnetize at room temperature and thus the layer does not affect the magnetic moment of the sample. The recorded MOKE signal also confirms that the ferromagnetic properties of the pure Ni layer does not suffer from an assumed oxidic passivation of the sample surface.

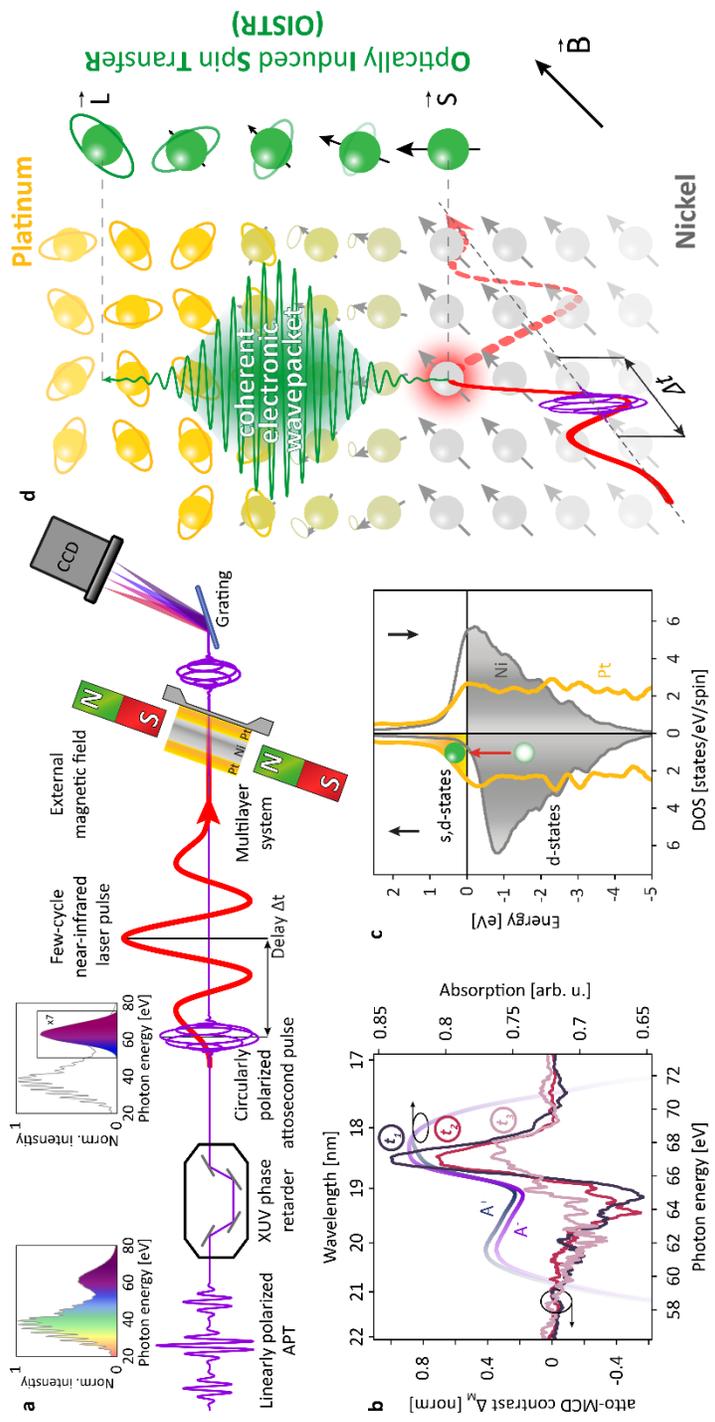

**Fig. 1: Atto-MCD to study coherent ultrafast magnetism**

A quarter wave phase retarder/band-pass filter optimized for photon energies covering the M-edge absorption of the most transition metals is used to turn incident linearly polarized attosecond pulse trains (APT) into isolated circular polarized attosecond pulses (panel **a**). These pulses are applied to record the magnetic circular dichroism contrast of nickel/platinum multilayer systems pumped by few-cycle near-infrared waveforms in an external magnetic field. Before the few-cycle near-infrared excitation, the nickel/platinum multilayer and the bulk nickel are in the magnetically saturated state along the in-plane direction easy axis set by $\pm$ ***B***.

Panel **b** displays the recorded absorption $A^+$ and $A^-$ for the two orientations of the magnetic field applied to the sample and the resulting atto-MCD contrast: before the arrival of the NIR pulse ($t_1$) which represents the unperturbed system, shortly after its arrival ($t_2$) and hundreds of femtoseconds after excitation ($t_3$).

In the multilayer (as opposed to bulk nickel) the optical excitation causes a coherent trans-interface charge current driven by the few-cycle light pulse, associated with which is the synchronous transfer of spin into the heavy metal (panel **c**), where the spin is subjected to the strong spin-orbit coupling of the heavy metal ad-layer (panel **d**).

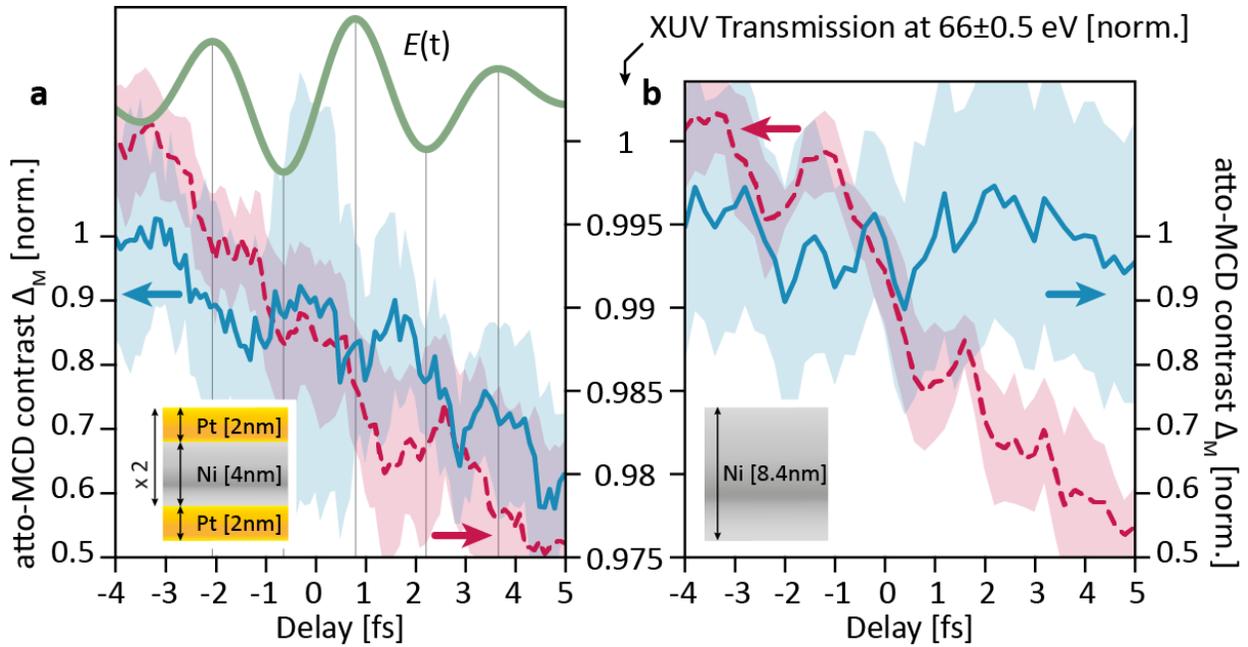

**Fig. 2: Attosecond-MCD of light-field induced coherent spin transfer**

The strong electric field $E(t)$ (waveform fitted to an attosecond streak camera recording, green line top of panel **a**) of an ultrashort laser pulse excites electrons around the Fermi energy in Ni/Pt-multilayer (panel **a**) samples and bulk nickel (panel **b**). The resulting transfer of electronic population in synchrony with the laser fields' half cycle oscillations is tracked with atto-XAS and yields a stepwise modification of the XUV transmission evaluated at $66 \pm 0.5$ eV (red-dashed curve in panel **a** and **b**). The simultaneously recorded atto-MCD contrast evaluated in the energy interval of $66.5 - 68$ eV, around the nickel $M_{2,3}$-edge is a measure of the magnetization of the nickel layer (blue solid curve in panel **a** and **b**). The reduction of the magnetic moment of the multilayer system synchronously with the electronic response is clear experimental evidence for ultrafast coherent spin transfer and the OISTR effect. This is evidenced further by bulk nickel showing no noticeable change in magnetic moment in the first femtoseconds after excitation (panel **b**), while also exhibiting an electronic response linked to the electric field oscillations of the excitation pulse.

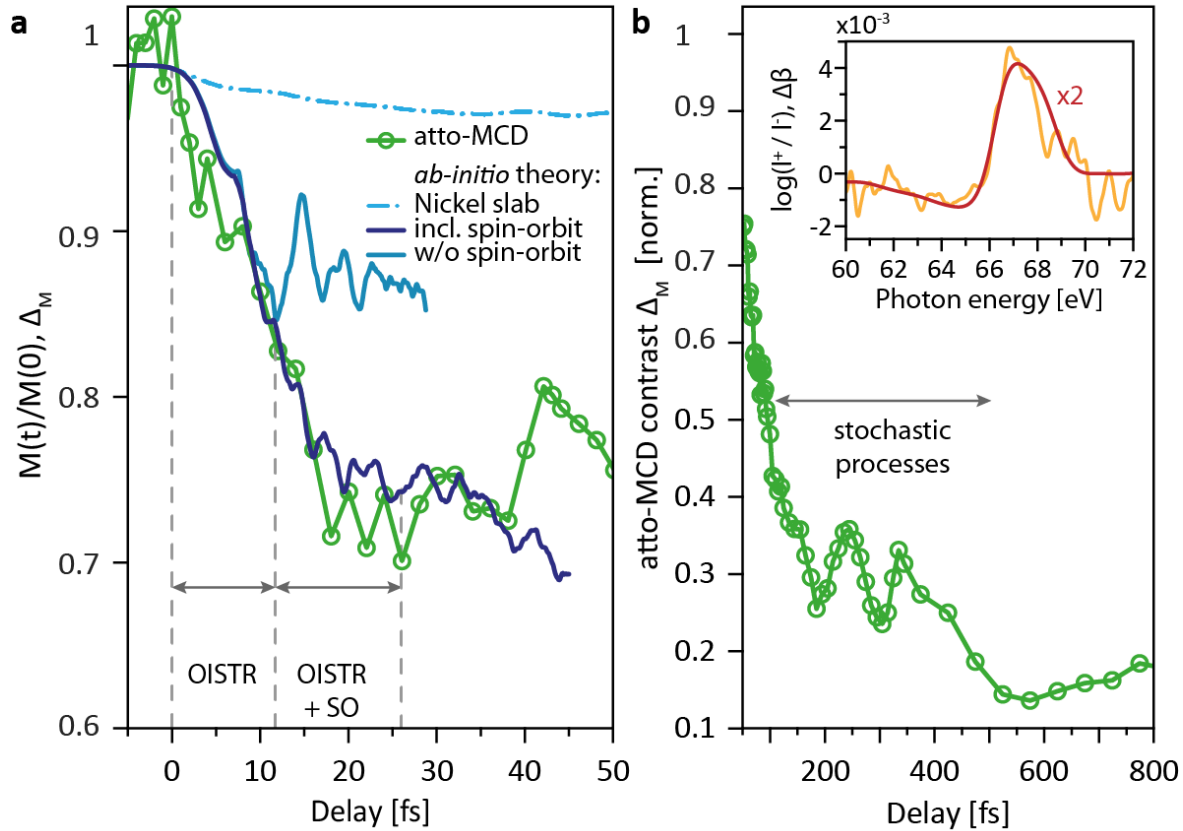

**Fig. 3: Optically induced spin-transfer, post-excitation spin dynamics and quantum dynamical modelling**

Comparison between the experimentally recorded atto-MCD trace (green line w. circles) in Ni/Pt-multilayers with the ab-initio simulation of light-field induced spin dynamics including spin-orbit coupling (dark-blue line, panel a) reveals that for the first 10 fs the demagnetization of Ni layers is entirely due to flow of majority spin current across the material interface. Theory without spin-orbit interaction predicts a saturation of the demagnetization already 15 fs after optical excitation (light-blue line). This is a clear indication that beyond this time all the demagnetization is caused by spin flips while at earlier times spin-dynamics are governed by OISTR – absent in the theoretical prediction for bulk nickel (dashed line and cp. Fig 2). Experimental data recorded for longer timescales (panel **b**) indicate the complete quenching of the magnetic moment due to stochastic processes with a characteristic timescale of 112 fs. The recorded $\Delta_M$ exhibits additional modulations up to 300 fs after optical excitation, that are potentially due to coherent phonon modes affecting the coupling to the lattice. The inset compares the recorded MCD signal $\Delta = \log \frac{I^+}{I^-}$ of the Ni/Pt sandwich with the imaginary part of the computed magneto-optical function.

**Extended Data**

Extended Data Figure 1: Hysteresis curves of the samples recorded using the longitudinal magneto-optical Kerr effect. Both samples exhibit soft magnetic hysteresis and low saturation fields needed to orient the macroscopic magnetization in the sample plane.

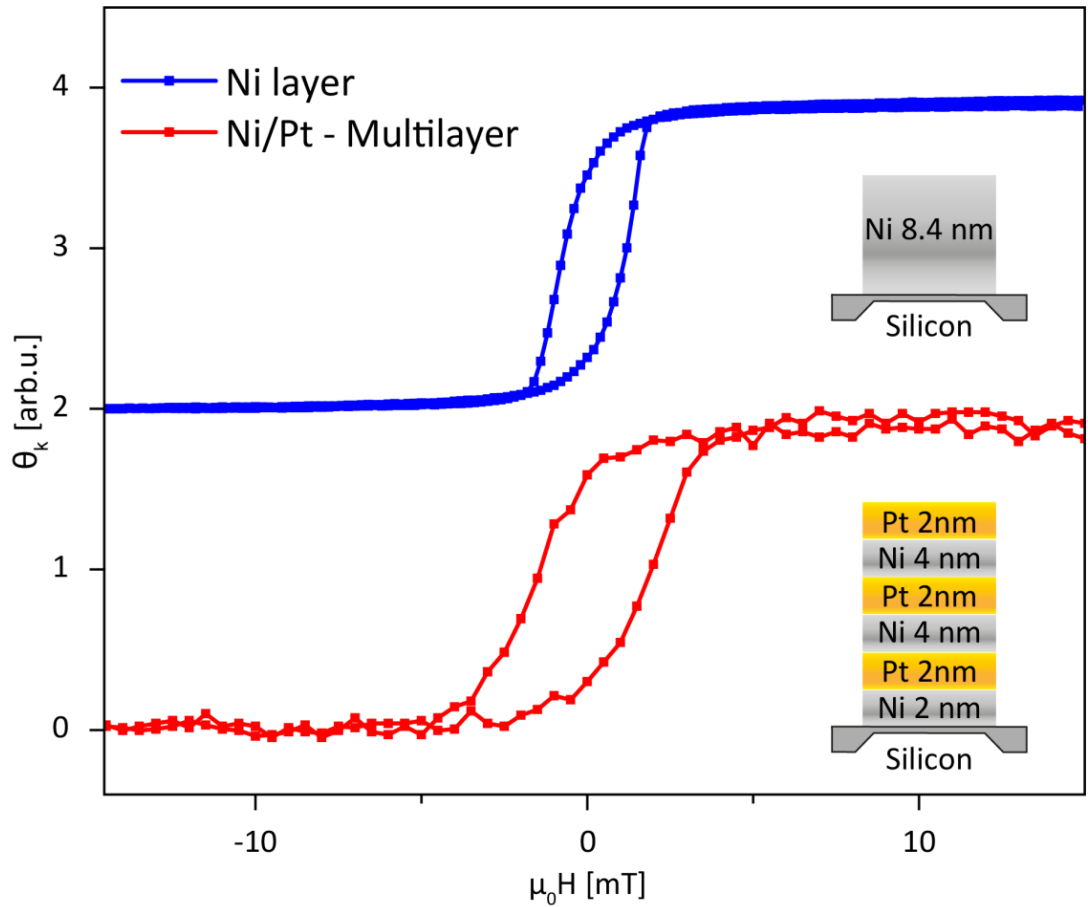